
\input aua.mac
\MAINTITLE{A Comparison of X-ray and Strong Lensing Properties of
Simulated X-ray Clusters}
\AUTHOR{Matthias Bartelmann\at{1} and Matthias Steinmetz\at{1,2}}
\INSTITUTE{\at{1}Max-Planck-Institut f\"ur Astrophysik,
Karl-Schwarzschild-Stra{\ss}e 1, D--85748 Garching, Germany;
\at{2}Present address: Department of Astronomy, University of
California, Berkeley, CA 94720, USA}
\ABSTRACT{We use gas-dynamical simulations of galaxy clusters to
compare their X-ray and strong lensing properties. Special emphasis is
laid on mass estimates. The cluster masses range between
$6\times10^{14}\,M_\odot$ and $4\times10^{15}\,M_\odot$, and they are
examined at redshifts between 1 and 0. We compute the X-ray emission
of the intracluster gas by thermal bremsstrahlung, add background
contamination, and mimic imaging and spectral observations with
current X-ray telescopes. Although the $\beta$ model routinely
provides excellent fits to the X-ray emission profiles, the derived
masses are typically biased low because of the restricted range of
radii within which the fit can be done. For $\beta$ values of
$\sim2/3$, which is the average in our numerically simulated sample,
the mass is typically underestimated by $\sim40$ per cent. The masses
of clusters which exhibit pronounced substructure are often
substantially underestimated. We suggest that the ratio between peak
temperature and emission-weighted average cluster temperature may
provide a good indicator for ongoing merging and, therefore, for
unreliable mass estimates. X-ray mass estimates are substantially
improved if we fit a King density profile rather than the $\beta$
model to the X-ray emission, thereby dropping the degree of freedom
associated with $\beta$. Clusters selected for their strong lensing
properties are typically dynamically more active than typical
clusters. Bulk flows in the intracluster gas contain a larger than
average fraction of the internal energy of the gas in such objects,
hence the measured gas temperatures are biased low. The bulk of the
optical depth for arc formation is contributed by clusters with
intermediate rather than high X-ray luminosity. Arcs occur
predominantly in clusters which exhibit substructure and are not in an
equilibrium state. Finally we explain why the X-ray emission of some
observed arc clusters is cooler than predicted from the arc
geometry. All clusters for which this happens in our simulations show
structure in velocity space, indicating ongoing merging along the
line-of-sight. The temperature discrepancy is probably a projection
effect.}
\KEYWORDS{Galaxies: clustering -- Cosmology: dark matter -- X-rays:
general}
%
%
%
\titlea{Introduction}
Until ten years ago, masses of galaxy clusters were mainly determined
using the kinematics of cluster galaxies. Although these mass
estimates are fairly crude (due mainly to small number statistics),
they already provided compelling evidence that most of the mass of
galaxy clusters is dark and is not concentrated in the cluster
galaxies themselves (Zwicky 1933). Within the last decade, this method
was complemented by two new, powerful methods based on the X-ray
emission of galaxy clusters and on their gravitational lens
effects. Similar to kinematic mass estimates, mass estimates based on
X-ray data generally rely on several {\it a priori} assumptions,
viz. (i) that the cluster mass distribution is spherically symmetric,
(ii) that the X-ray emitting gas is in hydrostatic equilibrium with
the dark-matter potential, and (iii) that the X-ray emitting gas is in
thermal equilibrium, in particular that its pressure is dominated by
thermal pressure. X-ray observations imply that the total mass of the
X-ray emitting gas is larger by about a factor five than that of the
cluster galaxies, but that the total baryonic mass from gas and
galaxies together is still smaller then the total cluster mass by
about a factor of five to ten.

Gravitational lensing of background galaxies opens a much more direct
way to infer cluster masses. Especially mass estimates based on weak
lensing, i.e. on the small image distortions of background galaxies,
appear very promising (e.g., Kaiser \& Squires 1993; Bonnet, Mellier,
\& Fort 1994; Seitz \& Schneider 1995; Broadhurst, Taylor, \& Peacock
1995; Squires \& Kaiser 1995; and references therein). It turns out,
however, that the power of weak lensing is more in obtaining the {\it
morphology} of the mass distribution within a galaxy cluster, while it
is much more difficult to deduce reliable estimates of the total mass
(e.g., Squires \& Kaiser 1995; Wilson, Cole, \& Frenk 1995; Seitz \&
Schneider 1996; Bartelmann et al. 1996; and references therein). In
addition, the essentially statistical analysis of weak-lensing effects
requires to average over background galaxy images and therefore has a
limited spatial resolution. Close to cluster centers, where most of
the X-rays are emitted, it is therefore difficult to produce
well-resolved mass maps. In addition, the contamination of the
background galaxy population by cluster galaxies is more severe close
to the cluster centers. Hence, while weak lensing can be expected to
yield important information about the outer parts of galaxy clusters,
strong lensing may provide more stringent constraints on their central
regions. At present, the masses inferred from these three methods,
viz. galaxy kinematics, X-ray emission, and gravitational lensing,
agree to within about a factor of two, but a systematic comparison
still needs to be done.

All observations performed so far demonstrate that the masses of
galaxy clusters are strongly dominated by dark matter. Although the
ratio between the total masses of baryonic and dark matter within
galaxy clusters is still a matter of debate (e.g., whether the hidden
mass in galaxy clusters is sufficient to account for a high-$\Omega_0$
universe), there is general agreement that this ratio is at least
five. From the theoretical point of view, such observations were
complemented by $N$-body and, in the more recent past, hydrodynamical
simulations (Thomas \& Couchman 1992; Schindler \& M\"uller 1993; Katz
\& White 1993; Cen \& Ostriker 1994; Bryan et al. 1994; Navarro,
Frenk, \& White 1995a; Evrard, Metzler, \& Navarro
1995). Gas-dynamical simulations performed so far have shown that
under most circumstances the mass estimates based on the assumptions
of spherical symmetry and hydrostatic equilibrium are reasonably
accurate, especially within an intermediate range of cluster
radii. Only in clusters which exhibit pronounced substructure is this
assumption obviously and demonstrably invalid. In addition, the
simulations have shown that the mass of clusters which are undergoing
a merging event is systematically underestimated because a fraction of
the energy of the gas is in the kinetic energy of bulk motions rather
than in thermal energy (Navarro et al. 1995a). Therefore the
temperature and consequently also the mass of such clusters are
underestimated. Nevertheless, Evrard et al. (1995) emphasize that for
clusters which do not clearly exhibit strong substructure, the mass
estimates are accurate to within 20 per cent and do not show a
systematic bias. They also argue that the accuracy of mass estimates
can be further improved using scaling relations. Their analysis was
however done under idealizing conditions in so far as (1) the
emission-weighted temperature of the simulated clusters was taken from
the simulations rather than from the cluster spectra, (2) the X-ray
images were analyzed out to large radii, and (3) only clusters at low
redshift were considered. For comparison with results from
gravitational lensing, however, clusters at intermediate redshifts
need to be studied because of the low lensing efficiency of
low-redshift clusters.

It is so far unclear which kinds of X-ray emitting clusters are
selected by the strong lensing effect. In order to be able to produce
large arcs, clusters have to be critical to lensing, which means that
a certain combination of their surface-mass density and shear field
must exceed a critical limit. This criterion has often been
interpreted such that the lensing clusters have to be the most massive
and hence most luminous and hottest X-ray clusters. It turns out,
however, that some of the clusters which show the most spectacular
arcs such as Abell 370 are only moderate X-ray emitters, while other
more prominent X-ray sources such as Abell 2163 are only moderately
strong lenses. Moreover, recent numerical studies (Bartelmann,
Steinmetz, \& Weiss 1995) have shown that the probability for a
cluster to form large arcs is significantly larger if the cluster is
substructured because the tidal field of a cluster with substructure
is larger than that of a spherically symmetric cluster, and the number
of cusps which are crucial for the formation of large arcs in the
caustic curves of cluster lenses is significantly enhanced by
asymmetries in the lens. Substructured clusters are, however, expected
to be far from equilibrium, and this should affect the X-ray
properties of those particular lenses. Miralda-Escud\'e \& Babul
(1995) in a recent study have found that two of three clusters for
which lensing- and X-ray data were available should have a higher
temperature to explain their lensing properties than the measured
temperature of the X-ray gas. It is important for our understanding of
the internal structure of galaxy clusters to identify the reason for
such temperature discrepancies.

Apart from the intrinsic interest in masses of galaxy clusters,
cluster masses are also of considerable importance for cosmology. For
example, a systematic underestimation of cluster mass based on X-ray
observations would favor a higher $\Omega_0$, a higher amplitude of
the power spectrum on scales of several Mpc, and it would also render
the baryon catastrophe (White et al. 1993; White \& Fabian 1995) less
severe. Thus it is important to identify possible systematic errors in
the estimation of cluster masses.

The present paper compares the X-ray and lensing properties of
numerically simulated galaxy clusters. The outline of the paper is as
follows. In Sect. 2, we describe our numerical simulations and discuss
the properties of our cluster sample. We describe how the X-ray and
lensing properties of the cluster sample are simulated and how
``detector characteristics'' are implemented. Section 3 concentrates
on mass estimates as obtained by X-ray ``observations'' of the cluster
sample. Section 4 concentrates on lensing and the properties of X-ray
clusters selected for their lensing effect. In Sect. 5 we summarize
and discuss our main results and conclusions.
\titlea{Numerical Cluster Simulations}
\titleb{The Simulations}
The simulations were performed with the Smoothed Particle
Hydrodynamics (SPH) code GRAPESPH (Steinmetz 1996). The background
cosmogony is an $\Omega_0=1$, $H_0=50\,$km~s$^{-1}$ CDM model
normalized to $\sigma_8=1$. The baryon fraction is $\Omega_{\rm
b}=0.05$. Our choice of the background cosmogony should be interpreted
as a convenient way to create a realistic sample of galaxy clusters in
different dynamical states, as opposed to trying to assess the merits
and shortcomings of high-$\Omega_0$ CDM. As a matter of fact, most of
our conclusions below are primarily caused by the hierarchical
assembly of galaxy clusters, while the details of the cosmogony are
probably of minor importance.

In order to perform simulations with sufficiently high resolution, we
applied a multiple mass scheme as described by Katz \& White (1993),
Navarro et al. (1995a), and Bartelmann, Steinmetz, \& Weiss
(1995). Based on a coarsely grained $N$-body simulation in a cube of
side length $300\,$Mpc, the 13 most massive clusters are
selected. These clusters are re-simulated in high--resolution runs
incorporating the effects of gas dynamics. In contrast to Katz \&
White (1993), but similar to Evrard et al. (1995) and Navarro et
al. (1995a), we do not include radiative cooling processes, mainly
because the interplay between gas cooling, feedback processes due to
the formation of stars within galaxies, and the limited numerical
resolution is not satisfactorily understood. On the other hand,
numerical simulations (Evrard et al. 1995; White, Efstathiou, \& Frenk
1993) which were performed neglecting radiative cooling show a
qualitatively good correspondence between observed and numerically
simulated X-ray clusters, though some scaling relations like for
example the temperature--luminosity relation of observed clusters are
not reproduced (Navarro et al. 1995a). Only thermal sources of gas
pressure are taken into account; the possible influence of magnetic
pressure is neglected. Nevertheless, we think that numerically
simulated galaxy clusters are fair representatives of real clusters in
the sense that they reproduce typical amounts of structure and typical
varieties of dynamic states. They are a generalization of the commonly
used hydrostatic models taking into account asymmetry as well as
internal dynamics. However, in spite of imperfections of the modeling,
statistical analyses and scaling relations should be applied with
caution.

The 13 cluster models projected along the three independent spatial
directions at $\sim8-10$ time steps each between redshifts 1 and 0
yield 378 simulated cluster images.
\titleb{X-ray Emission}
The temperature and density of the SPH gas particles are interpolated
to an equidistant grid with the usual SPH procedure (Lucy 1977;
Gingold \& Monaghan 1977; Monaghan 1992). At a grid point $\vec r$,
the SPH-averaged density is given by the sum
$$
  \langle\rho(\vec r)\rangle = \sum_{i=1}^{N}m_i\,
  W(\vec r-\vec r_i;h_i)\;,
\eqno(1)$$
where $i$ is running over all $N$ particles. The positions, masses,
and smoothing lengths of the particles are $\vec r_i$, $m_i$, and
$h_i$, respectively, and $W(\vec r;h)$ is the SPH smoothing
kernel. Given the density, the temperature at the grid position $\vec
r$ is determined by
$$
  \langle T(\vec r)\rangle = \sum_{i=1}^{N}m_i\,
  {T(\vec r_i)\over\rho(\vec r_i)}\,W(\vec r-\vec r_i;h_i)\;,
\eqno(2)$$
where $\rho(\vec r_i)$ is the density of the SPH particle $i$.

Given temperature and density of the gas in the grid cells, the
bremsstrahlung emission of each cell can be computed. It is sufficient
for our purposes to include continuum emission only, hence the
emissivity is
$$\eqalign{
  j(T,n_{\rm e}) &= {4\,C_{\rm j}\over1+f}\,
  \left({kT\over{\rm keV}}\right)^{1/2}\,
  \left({n_{\rm e}\over{\rm cm}^{-3}}\right)^2\,\cr
  &\times
  \left[\exp\left(-{E_{\rm a}\over kT}\right) -
        \exp\left(-{E_{\rm b}\over kT}\right)\right]\,
  {{\rm erg}\over{\rm cm}^3\,{\rm s}}\;,\cr
}\eqno(3)$$
where $f$ is the hydrogen fraction by weight of the gas, $n_{\rm e}$
is its electron number density, and $E_{\rm a,b}$ are the lower and
upper bounds of the energy band in which the emission is observed. We
choose $f=0.75$. With the Gaunt factor set to unity,
$$
  C_{\rm j} = 2.42\times10^{-24}\;.
\eqno(4)$$
The number of photons emitted from a grid cell is
$$\eqalign{
  \gamma(T,n_{\rm e}) &= {4\,C_\gamma\over1+f}\,
  \left({kT\over{\rm keV}}\right)^{-1/2}\,
  \left({n_{\rm e}\over{\rm cm}^{-3}}\right)^2\,\cr
  &\times
  \left[{\rm E}_1\left(-{E_{\rm a}\over kT}\right) -
        {\rm E}_1\left(-{E_{\rm b}\over kT}\right)\right]\,
  {1\over{\rm cm}^3\,{\rm s}}\;,\cr
}\eqno(5)$$
where ${\rm E}_1(x)$ is the exponential integral of the first
kind. Here, the constant factor is
$$
  C_\gamma = 1.51\times10^{-15}\;.
\eqno(6)$$
The spectrum of the photons is determined by
$$\eqalign{
  d\gamma(T,n_{\rm e}) &= {4\,C_\gamma\over1+f}\,
  \left({kT\over{\rm keV}}\right)^{-1/2}\,
  \left({n_{\rm e}\over{\rm cm}^{-3}}\right)^2\,\cr
  &\times
  \exp\left(-{E\over kT}\right)\,{dE\over E}\,
  {1\over{\rm cm}^3\,{\rm s}}\;.\cr
}\eqno(7)$$
The bounds of the energy band, $E_{\rm a,b}$, must be redshifted to
the cluster redshift.

Equations (3), (5), and (7) can then be combined to compute the total
X-ray luminosity, the X-ray flux, the photon flux, and the photon
spectrum observed from the model cluster by a detector specified by
its area, its energy band, and its spatial and spectral
resolutions. All cluster properties such as their masses are later
determined based on photon tables only, giving for each observed
photon its position in the detector and its energy. In addition, we
add a random X-ray background of $3\times10^{-4}$ s$^{-1}$
arcmin$^{-2}$ to the photon tables. This is lower by about an order of
magnitude than the high background of the ROSAT {\it High-Resolution
Imager} (HRI; David et al. 1993), but corresponds approximately to the
average background of the ROSAT {\it Position-Sensitive Proportional
Counter} (PSPC; Snowden et al. 1995).

We choose an exposure time of 20 ksec, which we reduce for very bright
clusters such that the total number of received photons does not
exceed $10^5$. As an immediate consequence of the limited exposure
time, the number of observed photons decreases for distant or faint
clusters, and the background becomes relatively more influential for
such clusters. We adopt a detector area of $10^4$ cm$^2$, similar to
the ROSAT telescope.

In order to mimic current X-ray telescopes, we choose spatial and
energy resolutions modeled after the ROSAT HRI (David et al. 1993) and
the ASCA {\it Solid-State Imaging Spectrometer} (SIS; Tanaka, Inoue,
\& Holt 1994). Images are taken with the properties of the HRI, which
has no energy resolution and a spatial resolution of a few arc
seconds. We assume a Gaussian point-spread function with $4''$
width. Spectra are taken with the energy resolution of the ASCA SIS,
i.e.,
$$
  {\Delta E\over E} = 0.02\,\left({E\over5.9\,
  {\rm keV}}\right)^{-1/2}\;.
\eqno(8)$$
We do not attempt to model the complicated ASCA point-spread function
in detail (Takahashi et al. 1995; Serlemitsos et al. 1995). Rather, we
take it to be Gaussian with $90''$ width. Consequently, the spatial
resolution of the simulated cluster spectra is much lower than that of
the synthetic cluster images. We choose this procedure such as to
model combined cluster analyses based on the best (i.e., most highly
resolved) cluster images and spectra available today. The HRI energy
band is $0.1\le E/{\rm keV}\le2.4$, the SIS band is $0.3\le E/{\rm
keV}\le12$.

\begfigside{fig1.tps}
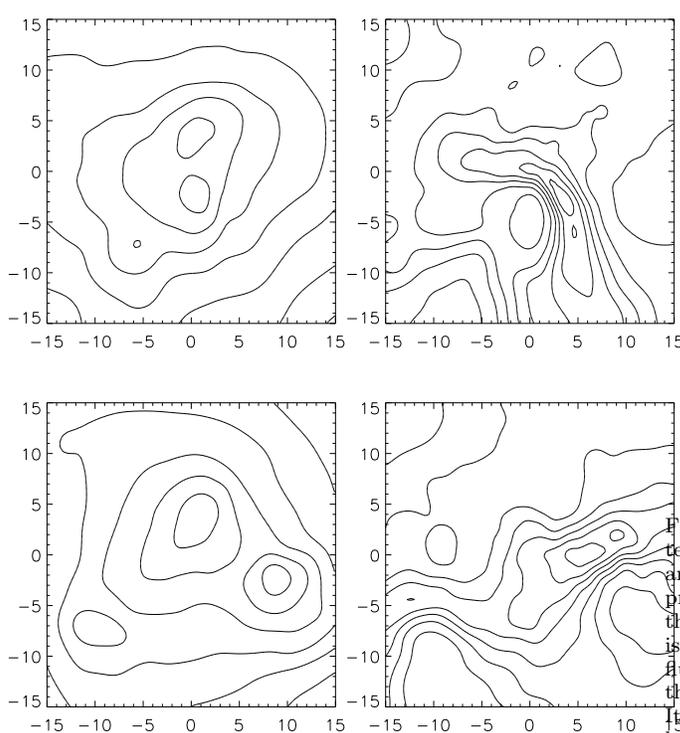
\figure{1}{Left column: Isodensity contours for a cluster which is
undergoing a merging event. The upper and lower panels correspond to
two different spatial projection directions. Left column: isoflux
contours of the X-ray emission; right column: (emission-weighted)
isotemperature contours of the intracluster gas. The flux contours are
logarithmically spaced by 0.5 decades, the temperature contours are
linearly spaced by 2 keV. It is seen in the top right frame that the
brighter clump has a bow shock, and the bottom right frame shows a
shock between the merging clumps.}
\endfig

\titleb{The Cluster Sample}
The cluster sample is virtually identical to that analyzed for its
lensing properties in Bartelmann et al. (1995), but it was
re-simulated including gas-dynamical effects. Within $r_{500}$ (the
radius of the sphere within which the averaged cluster density is 500
times the background density), the clusters have total masses between
$6\times10^{14}\,M_\odot$ and $4\times 10^{15}\,M_\odot$ and
line-of-sight velocity dispersions between 700 and 1200 km
s$^{-1}$. The mass resolution is $3\times10^{11}\,M_\odot$ in dark
matter and $1.6\times10^{10}\,M_\odot$ in gas, i.e. each cluster is
resolved into between 2000 and 12000 particles of each species. The
emission-weighted central temperatures range between $4\,$keV and
$13\,$keV, respectively. Many of the clusters are quite relaxed and
exhibit only small amounts of substructure. Their temperature profiles
are fairly flat; the central temperatures deviate from the averaged
emission-weighted temperature by less than a factor of about 1.5. The
isodensity and isotemperature contours are usually almost concentric.
Typically for a high-$\Omega_0$ cosmogony, the clusters undergo strong
evolution even at redshifts close to zero. However, due to our high
normalization ($\sigma_8=1$), our galaxy clusters are not as rare as
in the canonical CDM cosmogony normalized to the observed cluster
abundance (White et al. 1993) which corresponds to $\sigma_8=0.6$. The
amount of substructure in our cluster sample should, therefore,
provide a compromise between the extremes of a low-normalized
$\Omega_0=1$ and a high-normalized $\Omega_0=0.3$ universe. During an
epoch of merging the clusters exhibit bimodal or even more complex
density structures. Isodensity and isotemperature contours then
deviate significantly from each other (cf. Fig.~1). The temperature
maximum does not coincide with the density maximum but falls between
two merging subclumps or on the bow shock of an infalling clump
(Fig.~1, right column). Shocks are then abundant in the intracluster
gas, and there the peak gas temperatures $T_{\rm peak}$ are
substantially higher than the broad-band average temperatures $T_{\rm
avg}$.

\begfigside{fig2.tps}
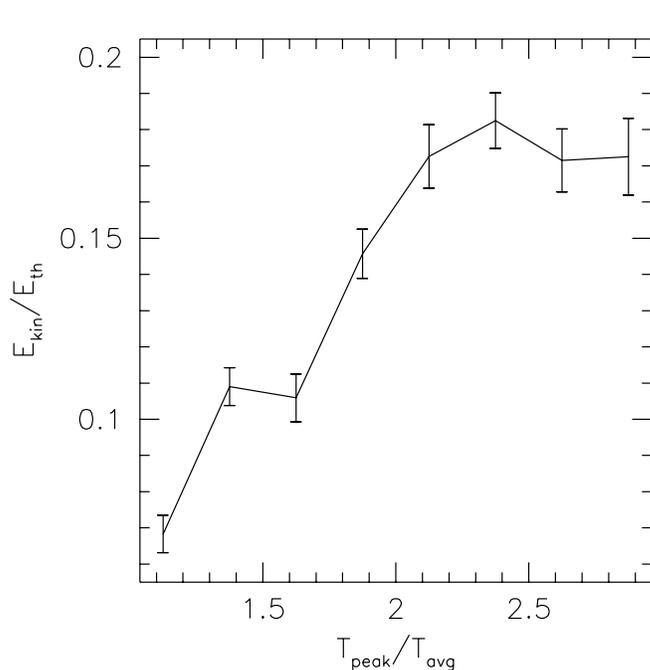
\figure{2}{Energy ratio $\epsilon$ as defined in eq.~(9) as a function
of the temperature excess $\tau$ (eq.~(10)). Unrelaxed clusters whose
kinetic energy is to a significant part in form of bulk motions of the
gas also exhibit a large ratio of maximum to averaged temperature,
i.e., the ratio $\tau$ is a good indicator for ongoing merging.}
\endfig

In Fig.~2 we plot the quantity
$$
  \epsilon\equiv{E_{\rm kin}\over E_{\rm th}}
\eqno(9)$$
as a function of
$$
  \tau\equiv{T_{\rm peak}\over T_{\rm avg}}\;,
\eqno(10)$$
where $E_{\rm kin,th}$ are the kinetic and thermal energies of the
intracluster gas, respectively. Values of $\epsilon$ close to zero
indicate complete thermalization of the gas, and the influence of bulk
motions increases with increasing $\epsilon$. As can be seen in
Fig.~2, those clusters which have a high $\tau$ also have high
$\epsilon$, i.e., they are not relaxed, and the kinetic energy of the
gas can reach 20 per cent of the thermal energy. Navarro et
al. (1995a) argued that ongoing merging is accompanied by a boost in
the velocity dispersion and, therefore, the ratio $\beta_{\rm
T}\equiv\mu\,m_{\rm p}\sigma_{\rm DM}^2/kT>1$ may be a good indicator
for merging.  For similar reasons, merging also causes a boost in the
peak temperature of the cluster ($\tau>2-3$). However, the ratio
$\tau$ is probably much more easily accessible to observation
(cf. Briel \& Henry 1994; Henry \& Briel 1995).
\titlea{Mass Determinations}
Given the photon tables for each cluster, we determine spectral
temperatures and then proceed to apply two techniques to determine
their masses.
\titleb{Temperature Determination}
Equation (7) yields the number of photons $d\gamma(T,n_{\rm e})$ per
energy interval $dE$ of a volume element of the intracluster gas with
temperature $T$ and electron density $n_{\rm e}$. Summing over all
volume elements (grid cells) and binning the photons according to the
energy resolution (8), we can compute synthetic spectra. For each
cluster image, such spectra are determined within rings of width $1'$
around the X-ray centroid, taking into account that the spatial photon
distribution is broadened by the PSF of the X-ray telescope.

Spectral temperatures are then found by fitting eq.~(7) to the
spectra. We find that the spectral temperatures are on average
accurate to $\sim10\ldots15$ per cent. The average 1--$\sigma$ error
of the best-fit temperatures is $\sim20$ per cent. Since X-ray cluster
mass estimates are proportional to the gas temperature (see below),
this scatter is directly propagated to an additional uncertainty in
the cluster masses. Within the error bars, there is no systematic
trend with temperature in the accuracy of the temperature
determination. Note, however, that the accuracy of our temperature
estimates is still idealized in so far as in the case of observed
clusters the accuracy can be further reduced by imperfections in the
discrimination between bremsstrahlung emission and emission in
individual metal lines, which depends on, e.g., the assumptions about
metal abundances in the intracluster gas.
\titleb{$\beta$-fit Models}
The most common technique to determine cluster masses from their X-ray
emission is the $\beta$-fit method, pioneered by Cavaliere \&
Fusco-Femiano (1976) and widely used thereafter (e.g., Jones \& Forman
1984; Henriksen \& Mushotzky 1985; Edge \& Stewart 1991). The photon
flux is azimuthally averaged in radial bins around the observed X-ray
centroid, yielding a flux profile of the cluster. The three-parameter
function
$$
  S_{\rm X} = S_0\,\left(1+x^2\right)^{-3\beta+1/2}
\eqno(11)$$
is then fit to the observed profile, where $x$ is the distance from
the X-ray centroid in units of a core radius $r_{\rm c}$. The three
parameters $S_0$, $r_{\rm c}$, and $\beta$ are then adjusted such as
to optimize the fit. $\beta$-fits obtained this way usually provide
excellent fits to the X-ray emission profile, both for real and
for simulated cluster emission (cf. the example profile and fit in
Fig.~4).

We can ignore the weak temperature dependence $\propto T^{1/2}$ of
thermal bremsstrahlung emission, because the intracluster gas is
almost isothermal over the radial range of the fit. The X-ray emission
is then determined by the square of the gas density, and hence the
form of the fit in eq.~(11) implies
$$
  \rho_{\rm gas}(r) = \rho_{\rm gas,0}\,(1+x^2)^{-3\beta/2}\;.
\eqno(12)$$

If the gas is in hydrostatic equilibrium in a spherically symmetric
potential well produced by the dark-matter distribution, and if it is
isothermal, then Euler's equation implies
$$
  M(r) = -{r\,kT\over G\mu m_{\rm p}}\,
  {d\ln\rho_{\rm gas}\over d\ln r}\;,
\eqno(13)$$
where $G$ is Newton's constant, $m_{\rm p}$ is the proton mass, and
$$
  \mu = {4\over5f+3}\;,\quad\mu=0.59\quad\hbox{\rm for}\quad f=0.75
\eqno(14)$$
is the mean molecular weight of a hydrogen-helium mixture with
hydrogen fraction $f$ (by weight). Given $\beta$ and $r_{\rm c}$ from
fitting the X-ray surface brightness, eqs. (12) and (13) imply
$$
  M(r) = {3\beta r\,kT\over G\mu m_{\rm p}}\,
  {x^2\over1+x^2}\;.
\eqno(15)$$
Figure 3 shows a histogram of best-fit $\beta$ values obtained from
our cluster sample. The distribution of values obtained from our
simulations compares well with observationally determined
distributions (e.g., Jones \& Forman 1984; Henriksen \& Mushotzky
1985; Edge \& Stewart 1991).

\begfigside{fig3.tps}
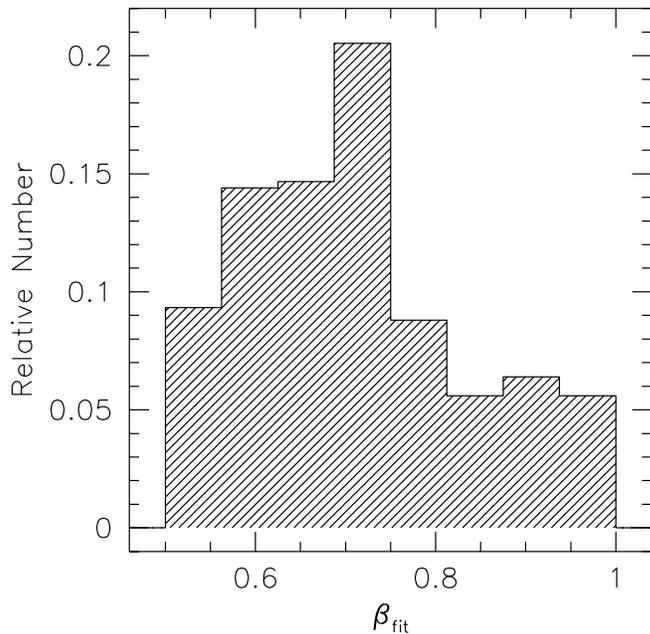
\figure{3}{Distribution of $\beta$ values fit to the cluster X-ray
profiles. The average is close to $\langle\beta\rangle\sim2/3$.}
\endfig

Note that eq.~(15) implies the dark-matter density profile
$$
  \rho(r) =
  \rho_0\;{3+x^2\over3(1+x^2)^2}\;,\quad
  \rho_0 = \rho(0) = {9\beta kT\over4\pi G\mu m_{\rm p}r_{\rm c}^2}\;.
\eqno(16)$$
The density profile asymptotically approaches the isothermal fall-off
$\propto x^{-2}$. The density profiles of our numerically simulated
clusters however are of the form recently found by Navarro, Frenk, \&
White (1995b),
$$
  \rho(y) = {\rho_{\rm s}\over y(1+y)^2}\;,
\eqno(17)$$
where $y$ is the radius in units of a scaling radius $r_{\rm
s}$. Since this profile is steeper by one power of $r$ than the
isothermal $\beta$-fit density profile, the isothermal $\beta$-fit
mass profiles are flatter than the true profiles. In particular, the
accuracy of $\beta$-fit mass estimates depends on radius.

For gas in hydrostatic equilibrium with the dark-matter potential
which has the same temperature as the dark-matter particles,
$$
  kT = \mu m_{\rm p}\,\sigma_\parallel^2\;,
\eqno(18)$$
where $\sigma_\parallel$ is the line-of-sight velocity dispersion of
the dark-matter particles. The physical meaning of the parameter
$\beta$ in hydrostatic equilibrium is
$$
  \beta = {\mu m_{\rm p}\,\sigma_\parallel^2\over kT}\;.
\eqno(19)$$
Since eq.~(18) is well fulfilled in our numerical simulations, and
since the simulated dark-matter density profiles asymptotically fall
off $\propto r^{-3}$, we should expect to obtain $\beta\sim1$ from
fitting the X-ray profiles with the $\beta$ model (11). The reason why
we find systematically low $\beta$ values is illustrated in
Fig.~4. There, we show the same background-subtracted X-ray profile
four times, each time with a different level for the adopted X-ray
background which is indicated by the dotted horizontal lines. The
$\beta$ model is fit to the X-ray emission profile out to the radius
where the profile falls below the background level. The secondary peak
at $r\sim10'$ is excluded from the fit. The solid line is the best-fit
$\beta$ profile obtained, and the optimal $\beta$ values are given in
each frame.

\begfigside{fig8.tps}
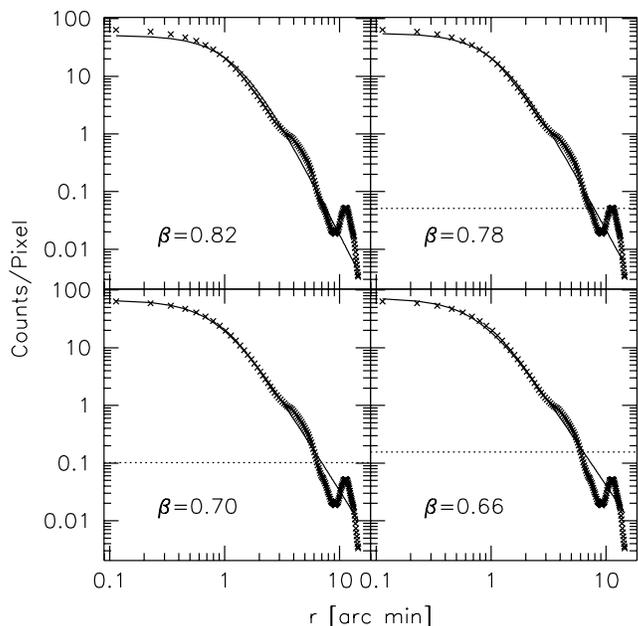
\figure{4}{The same simulated X-ray emission profile is plotted four
times (crosses). The adopted X-ray background is increased from zero
(top left frame) to $[1,2,3]\times10^{-4}$ s$^{-1}$ arcmin$^{-2}$ (top
right, bottom left, and bottom right frame, respectively) as indicated
by the dotted horizontal lines. The solid lines in all frames are the
$\beta$ models obtained by fitting the profile out to the radius where
they fall below the background. With increasing background intensity
and thus decreasing radial range of the fit, $\beta$ drops from $0.82$
to $0.66$. The secondary peak at $r\sim10'$ is excluded from the fit.}
\endfig

The figure shows that the best-fit $\beta$ value decreases while the
background is gradually increased. Without the background,
$\beta=0.82$ (top left frame), and when the background reaches the
level of $3\times10^{-4}$ s$^{-1}$ arcmin$^{-2}$ adopted throughout
our simulations, $\beta$ drops to $0.66$. Hence, the reason for the
systematically low $\beta$ values in our simulations is the limited
range of radii accessible with X-ray observations. Where the emission
reaches the background level, the profile has not yet reached its
asymptotic slope.

The observed discrepancy between $\beta_{\rm fit}$ as obtained from
fitting the profile (11) and $\beta_{\rm spec}$ as obtained through
(19) from measuring $T$ and $\sigma_\parallel$ has been widely
discussed in the literature (see, e.g., Sarazin 1986 for a
review). Since our numerically simulated clusters satisfy $\beta_{\rm
spec}\simeq1$, they reproduce the observed $\beta$
discrepancy. Several solutions to the $\beta$ problem have been
suggested. Lubin \& Bahcall (1993) and Bahcall \& Lubin (1994) showed
that the $\beta$ discrepancy disappeared when they adopted the actual
radial profile of the cluster galaxy distribution rather than the King
profile underlying eq. (11), which is equivalent to assuming a flatter
dark-matter density profile. Evrard (1990) ascribed the $\beta$
discrepancy in his numerically simulated galaxy clusters to a
combination of incomplete thermalization of the intracluster gas and
anisotropic dark-matter velocity distributions. Navarro et al. (1995a)
also found the $\beta$ discrepancy in their models and concluded that
(1) $\beta_{\rm fit}$ is biased low because of the limited radial
range available for fitting the X-ray profile and (2) the actual slope
of the dark matter profile in the relevant radial range is smaller
than its asymptotic slope. Although the intracluster gas in our
simulations is not completely thermalized either, this has only a weak
influence on the $\beta$ discrepancy because on average the kinetic
energy of the gas is only $\sim10$ per cent of the thermal energy
(cf. Fig. 10 below). Our results therefore support the view of Navarro
et al. (1995a) that the limited radial range accessible to cluster
observations is sufficient to explain the $\beta$ discrepancy.

Figure 4 illustrates an additional effect which can cause $\beta$ fits
to be systematically low. At $r\sim4'$ in the example plotted there,
there is a low hump in the profile which is due to the emission of a
small infalling subclump. Despite this subclump, the cluster image
looks almost spherically symmetric. Although weak, the hump in the
profile flattens the best-fit $\beta$ model if it is close to the
background cutoff as in the case of the profile shown in Fig.~4.
\titleb{Masses Based on the King Profile}
We have seen that the $\beta$ model yields systematically low $\beta$
values because of the restricted radial range within which the fits
can be done. Since we have good theoretical reasons to believe that
dark-matter density profiles asymptotically fall off $\propto r^{-3}$
(Navarro et al. 1995b; Cole \& Lacey 1995), it appears preferable to
base X-ray mass estimates on the assumptions that (1) the dark-matter
density profile can be described by some profile $\propto r^{-3}$ for
large $r$ and (2) that the X-ray gas is in thermal and hydrostatic
equilibrium with the dark-matter potential. Then, the gas density
profile has the same shape as the dark-matter density profile. We
should however add the cautionary note that it has not been verified
yet that actual galaxy clusters follow the density profiles found
numerically.

To illustrate that procedure, we choose King's approximation to the
isothermal profile (King 1966; Binney \& Tremaine 1987),
$$
  \rho(r) = \rho_0\,(1+y^2)^{-3/2}\;,
\eqno(20)$$
where $y$ is the distance from the X-ray centroid in units of the King
radius,
$$
  r_{\rm K} = \left({9\sigma_\parallel^2\over
  4\pi G\rho_0}\right)^{1/2}\;. 
\eqno(21)$$
The integrated dark mass can then be written
$$\eqalign{
  M(r) &= {9kT\,r_{\rm K}\over G\,\mu m_{\rm p}}\,f(y)\;,\cr
  f(y) &\equiv \ln\left(y+\sqrt{1+y^2}\right) -
  {y\over\sqrt{1+y^2}}\;.\cr
}\eqno(22)$$

We find the King radius $r_{\rm K}$ in our simulations by fitting the
profile
$$
  S_{\rm X}' = S_0'\,(1+y^2)^{-5/2}
\eqno(23)$$
to the simulated X-ray profiles. This is equivalent to fitting a
$\beta$ model under the constraint $\beta=1$. We note that this is
similar to the approach suggested by Evrard et al. (1995) which
consists in constraining the value of $\beta$ with scaling laws
obtained from numerical simulations.
\titleb{Results for Mass Estimates}
We now display and summarize the results for the two different methods
to estimate X-ray masses which we have applied to the numerically
simulated clusters. Figure 5 shows the ratio between estimated and
true cluster mass as a function of radius. The radii are scaled by the
radius $r_{500}$ within which the density contrast of the cluster
relative to the background density is 500 in order to superpose the
results obtained from all simulated clusters. We restrict the plots to
clusters at redshifts below $0.5$. The frames in the left column of
Fig.~5 show results for low-redshift clusters ($0\le z<0.25$), and the
results for intermediate-redshift clusters ($0.25\le z<0.5$) are
displayed in the right column. The top frames show $\beta$-fit mass
estimates, and the bottom frames show mass estimates assuming a King
profile for both the dark-matter and the gas density. The error bars
give the 1--$\sigma$ scatter in all applicable simulations.

\begfigside{fig4.tps}
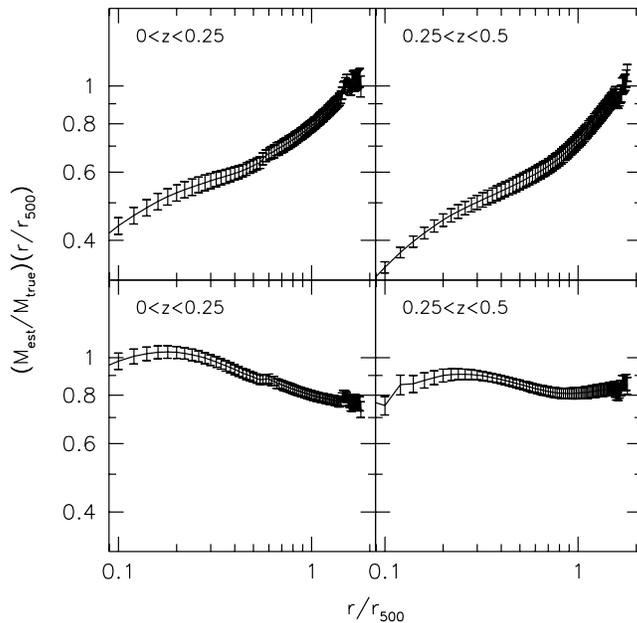
\figure{5}{Ratio between the estimated and the true cluster mass
$M_{\rm est}/M_{\rm true}$ as a function of radius. The radii are
scaled by the radius $r_{500}$ so that the results from all clusters
can be superposed. The left (right) frames are for low-
(intermediate-) redshift clusters as indicated, the top (bottom)
frames are for $\beta$ (King-profile) fits, respectively. While the
mass ratio obtained from $\beta$ fits rises with $r$, it is almost
flat for the King-profile fits. This reflects the fact that the
isothermal mass profiles obtained from the $\beta$ fits are shallower
than the true mass profiles. Error bars indicate the 1--$\sigma$
scatter over all applicable simulations.}
\endfig

First, we note that there is no substantial difference between the
low- and the intermediate-redshift cluster samples. The accuracy of
the $\beta$-fit mass estimates depends on radius as discussed before:
Since the isothermal $\beta$-fit mass profile is shallower than the
true mass profile, the ratio between the estimated and the true mass
increases with radius. In contrast to that, the estimates assuming the
King profile are almost flat within the error bars, and over most of
the range of radii accessible to X-ray observations the accuracy is
significantly improved over that of the $\beta$-fit mass
estimates. The slightly falling curves in the bottom left frame
indicate that for the low-redshift cluster sample the King profile is
slightly too steep. This is due to merging of subclumps which renders
radially averaged mass profiles shallower than the King profile
expected for an isolated cluster. Also note that the mass estimates
from the King model are biased somewhat low. We will discuss the
reason below.

Some further features of the different mass estimates are illustrated
in Figs. (6) and (7). Figure 6 displays histograms for the
distribution of cluster mass estimates at the maximum radius $r_{\rm
obs}$ accessible to X-ray observations, where the surface brightness
drops to the background level. The frames are arranged as in Fig.~5:
The top frames show $\beta$-fit results, the bottom frames show
results from the assumption of a King profile. The left and right
columns displays results for low- and intermediate-redshift cluster
samples, respectively.

\begfigside{fig6.tps}
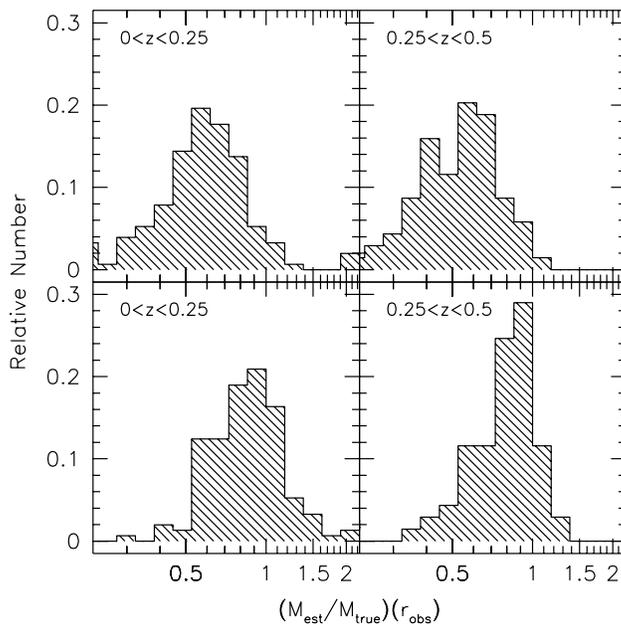
\figure{6}{Histograms for the distribution of the ratio $M_{\rm
est}/M_{\rm true}$ at the maximum observable radius $r_{\rm obs}$. The
frames are arranged as in Fig.~5. The distributions of the $\beta$-fit
mass estimates peak at $\sim0.6$, while the estimates from
King-profile fits peak at $\sim0.9$. The scatter in the $\beta$-fit
results is much larger than for the King-profile fits because of the
uncertainty in the best-fit values of $\beta$.}
\endfig

The $\beta$-fit mass estimates are centered on $M_{\rm est}/M_{\rm
true}\sim0.6$, while the King-profile estimates peak at
$\sim0.9$. Also, the width of the distributions is reduced by fitting
the King profile because the scatter in the best-fit $\beta$ values is
removed.

To further highlight the influence of the systematic bias in the
$\beta$ values on the accuracy of the mass estimates, Fig.~7 displays
the ratio $M_{\rm est}/M_{\rm true}$ at the observable radius $r_{\rm
obs}$ as a function of $\beta$.

\begfigside{fig5.tps}
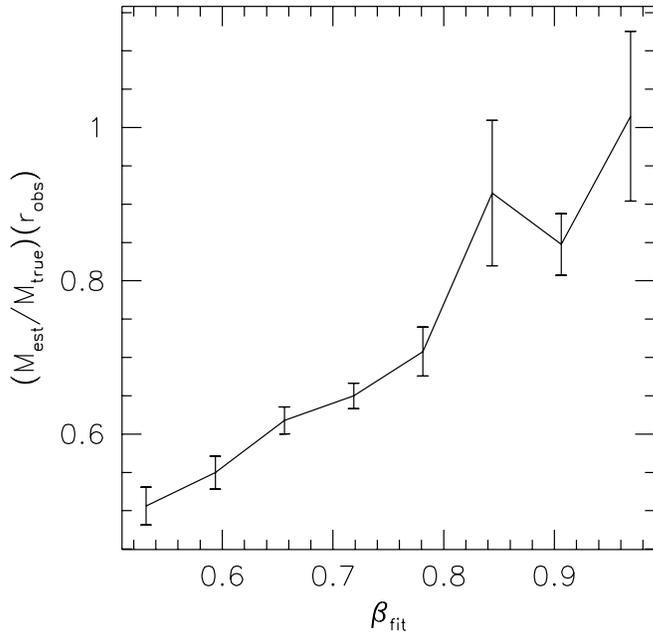
\figure{7}{Ratio $M_{\rm est}/M_{\rm true}$ from $\beta$ fits at the
radius $r_{\rm obs}$ as a function of $\beta$. The curve shows a
roughly linear relation between the accuracy of the mass estimates and
$\beta$. For $\beta\sim1$, the masses estimates are accurate, and they
drop to $\sim60$ per cent of the true mass for the average value
$\beta\sim2/3$.}
\endfig

As expected from eq.~(15), Fig.~7 shows a roughly linear relation
between the accuracy of the mass estimates and $\beta$. For clusters
with $\beta\sim1$, the mass estimates are almost exact, and they drop
in proportion to $\beta$. At the average $\beta\sim2/3$, the estimated
mass at $r_{\rm obs}$ is $\sim60$ per cent of the true mass.

As mentioned before, also the mass estimates based on the King-profile
fits are systematically low by $\sim10$ per cent (e.g., the histograms
in the bottom frames of Fig.~6 peak at $M_{\rm est}/M_{\rm
true}\sim0.9$). The reason for that is illustrated in Fig.~8, where we
plot as the solid line the accuracy of the mass estimates from the
King profile in dependence on the ratio between the kinetic and the
thermal energy of the gas. The dotted line shows the mass estimates
multiplied by the factor $1+E_{\rm kin}/E_{\rm th}$.

\begfigside{fig7.tps}
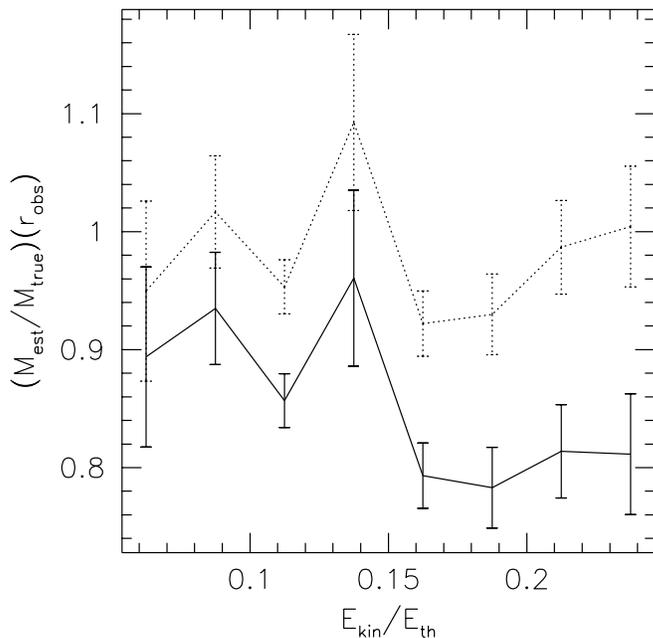
\figure{8}{Solid line: Ratio $M_{\rm est}/M_{\rm true}$ obtained from
King-profile fits at the radius $r_{\rm obs}$ as a function of the
ratio $E_{\rm kin}/E_{\rm th}$ between the kinetic and the thermal
energy of the cluster gas. The dotted line shows the same results
multiplied by the factor $1+E_{\rm kin}/E_{\rm th}$ by which the gas
temperature should be higher if the gas was fully thermalized. The
figure shows that mass estimates are biased low in the presence of
bulk flows in the cluster gas because then the gas temperature is
systematically low. The corrected curve shows that the mass estimates
are accurate to within the error bars if the cluster gas is completely
thermalized.}
\endfig

The figure shows that the accuracy of the mass estimates decreases
with increasing fraction of kinetic relative to thermal gas
energy. This fraction indicates the presence of bulk flows in the
gas. In the presence of bulk flows, the gas is not fully thermalized,
and therefore the gas temperature is lower than it should be if
pressure equilibrium was maintained by thermal gas pressure only. Low
temperature biases the mass estimates low. If the gas can be
considered in virial equilibrium with the cluster potential but not
yet in thermal equilibrium, this bias should amount to the ratio
between the thermal energy and the sum of thermal and kinetic energy
because the mass is proportional to the gas temperature. Therefore, if
we multiply the mass estimates by the factor $1+E_{\rm kin}/E_{\rm
th}$, we correct for the low temperature. Although the scatter is
large, the corrected (dotted) curve in Fig.~8 shows that the corrected
mass estimates are accurate to within the error bars.
\titlea{Properties of Lensing X-ray Clusters}
\titleb{Arc Cross Sections}
Our method to investigate the arc-formation statistics of the
numerical cluster models was described in detail by Bartelmann \&
Weiss (1994). We will therefore keep the present description brief and
refer the reader to that paper for further information. For general
information on gravitational lensing, see Schneider, Ehlers, \& Falco
(1992) and references therein.

The numerical cluster models yield the spatial coordinates and
velocities of discrete particles with equal mass. In order to use them
for gravitational lensing, we need to compute the surface mass density
distribution of each cluster model in each of the three independent
directions of projection. The mass density is first determined on a
three-dimensional grid according to eq.~(1) and subsequently smoothed
with a Gaussian filter function. The grid resolution and the width of
the Gaussian are adapted to the numerical resolution of the code in
order not to lose spatial resolution by the smoothing of the density
field. The smoothed density field is then projected onto the three
sides of the computation volume to obtain three surface-density fields
for each cluster.

The physical surface mass density fields are then scaled by the
critical surface mass density for lensing, which apart from the
cosmological parameters depends on the cluster- and source
redshifts. We keep the redshift for all sources fixed at $z_{\rm
s}=1$, and the cluster redshifts vary between $0\le z_{\rm
c}\le1$. This finally yields three two-dimensional convergence fields
$\kappa(\vec x)$ at the two-dimensional position $\vec x$ for each
cluster model. From $\kappa(\vec x)$, all quantities determining the
local lens mapping, i.e., the deflection angle and its spatial
derivatives, can be computed. We determine the lens properties of the
clusters on grids with an angular resolution of $0\farcs3$ in the lens
plane in order to ensure that lensed images be properly resolved.

Sources are then distributed on a regular grid in the source
plane. The resolution of this source grid can be kept low close to the
field boundaries because there no large arcs occur. Close to the
caustics of the clusters, where the large arcs are formed, the
source-grid resolution is increased with the increasing strength of
the lens. For our later purpose of statistics, sources are weighted
with the inverse resolution of the grid on which they are placed. The
sources are taken to be intrinsically randomly oriented ellipses with
their axis ratios drawn randomly from the interval $[0.5,1]$, and
their axes determined such that their area equals that of circles with
radius $0\farcs5$. Although this choice of source properties appears
fairly simple, it should not affect the arc statistics because the
latter mainly reflects the local properties of the lens mapping, which
are independent of the particular choice of source sizes or the
ellipticity distribution.

The sources are then viewed through the cluster lenses. All images are
then classified in the way detailed by Bartelmann \& Weiss
(1994). Among other things, the classification yields for each image
its length $L$ and its width $W$. In total, we classify the images of
about 78,000 sources lensed by 80 cluster models.

Knowing the area covered by the cluster fields, and having determined
the frequency of occurrence of image properties such as a given length
and width, we can compute cross sections for the formation of images
with such properties. This procedure yields tables $\sigma_{ijk}$ of
cross sections in units of ($h^{-1}\,$Mpc)$^2$, one each for cluster
model $i$, projected along the spatial direction $j$, at time or
redshift $k$. We then interpolate these cross-section tables between
the discrete redshifts to obtain cross sections $\sigma_{ij}(z)$ as
smooth functions of cluster redshift $z$ for each cluster $i$ and each
projection direction $j$, and finally average these cross sections
over all $i$ and $j$. The result is the average cross section per
cluster $\langle\sigma\rangle(z)$ as a function of redshift. Examples
for the cross sections $\langle\sigma\rangle(z)$ can be found in
Bartelmann et al. (1995).
\titleb{Optical-Depth Weighted Averages}
We want to study properties of such X-ray clusters which are selected
for their lensing effects, more specifically for their ability to
produce arcs. We therefore need to introduce an averaging procedure
which weights the cluster properties under consideration with the arc
optical depth contributed by the individual clusters. A
straightforward way to do so (cf. Bartelmann 1995) is to define the
optical-depth weighted average of a quantity by
$$
  \bar{Q}_{ij} = {1\over\tau_{ij}(z_{\rm s})}\int_0^{z_{\rm s}}dz\,
  {d\tau_{ij}(z)\over dz}\,Q_{ij}(z)\;,
\eqno(24)$$
where the model indices $i$ and $j$ indicate that the averages are to
be taken for each cluster model and each projection direction
separately. The optical depth $\tau_{ij}(z_{\rm s})$ is the optical
depth for lensing of sources at redshift $z_{\rm s}$, contributed by
the cluster model specified by $(i,j)$, which is defined as the
probability with which a source at redshift $z_{\rm s}$ is imaged as
an arc with certain specified properties. It is given as an integral
over lens redshift of the cross section $\sigma_{ij}(z)$ of the given
cluster model, weighted by the proper volume and divided by the area
of the source sphere,
$$
  \tau_{ij}(z_{\rm s}) = {1\over4\pi D_{\rm s}^2}
  \int_0^{z_{\rm s}}dz\,\left|{dV(z)\over dz}\right|\,
  n_0\,(1+z)^3\sigma_{ij}(z)\;.
\eqno(25)$$
$D_{\rm s}$ is the angular-diameter distance to the source, $dV(z)$ is
the proper volume element at redshift $z$, $n_0$ is the comoving
cluster density, and the factor $(1+z)^3$ accounts for the cosmic
expansion. The quantity $\bar{Q}$ averaged over the entire cluster
sample is finally given by
$$
  \bar{Q} = {1\over\tau(z_{\rm s})}\sum_{ij}\tau_{ij}(z_{\rm s})
  \bar{Q}_{ij}\;.
\eqno(26)$$
Of course, the results depend on the arc properties chosen. In the
following, we choose the arc length-to-width ratio $L/W\ge10$ to
perform the averages (24) and (26), i.e., we select for giant arcs.
\titleb{Arc-Cluster Properties}
We start by comparing different estimates of the total cluster mass
within the arc radius. We show in Fig.~9 the distribution of the true
cluster mass (i.e., the total mass of all cluster particles) in panel
(a) together with the distributions of mass estimates derived from
lensing [panel (b)], and from $\beta$- and King-profile fits to the
cluster X-ray emission [panels (c) and (d), respectively].

\begfigside{fig9.tps}
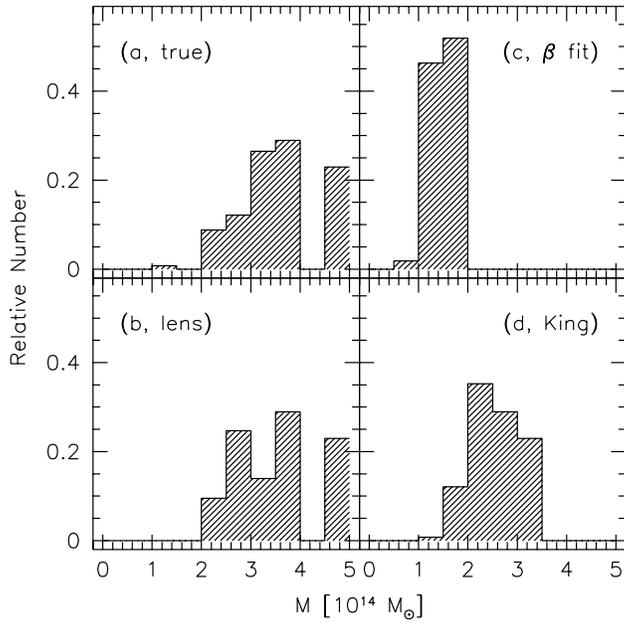
\figure{9}{Distributions of cluster masses in $10^{14}\,M_\odot$ and
different mass estimates at the arc radius. (a) true cluster mass; (b)
lensing mass, i.e., projected mass; (c) $\beta$-fit mass estimates;
(d) mass estimates from King-profile fits. The comparison between
frames (a) and (b) show that the distributions of lensing- and true
masses are almost identical, indicating that projection effects of
lensing clusters are statistically unimportant. The $\beta$ fit mass
estimates are systematically low by a factor of $\sim2-3$, while the
mass estimates from the King-profile fits are low by $\sim20$ per
cent.}
\endfig

The lensing mass estimate is given by the average convergence
$\bar\kappa$ within circles traced by the arcs, multiplied by the area
enclosed by the circles and the critical surface mass density
$\Sigma_{\rm cr}$. The lensing mass estimate is therefore the
projected cluster mass within a cylinder around the line-of-sight with
radius equal to the distance between the arc and the cluster
center. For spherically symmetric clusters, $\bar\kappa=1$ by
definition of the tangential critical curve. For asymmetric clusters,
$\bar\kappa$ is systematically below unity (Bartelmann 1995), hence
the simple mass estimate based on $\bar\kappa=1$ would overestimate
the projected cluster mass. The true $\bar\kappa$ which we adopt here
is not directly accessible through observations of strong lensing, but
the success of cluster mass models reproducing observed arcs and
multiple images indicates that reliable estimates of $\bar\kappa$ can
be given in practice.

A comparison between panels (a) and (b) of Fig.~9 shows that the
distributions of the true dark mass and the lensing mass agree very
well although the lensing mass is projected along the line of
sight. If lensing clusters were preferentially elongated along the
line-of-sight such as to increase their surface mass density, the
lensing mass would be significantly larger than the true mass. The
fact that this is not the case demonstrates that such a selection
effect on arc clusters is at least statistically unimportant. The
distribution of the $\beta$-fit mass estimates (panel c) peaks at
$\sim1.5\times10^{14}\,M_\odot$, while the distribution of true
cluster masses peaks at about $2.5$ times this value. This reflects
the fact that best-fit $\beta$ values are systematically too low
especially at the rather small radii where large arcs appear in
clusters, and that the isothermal dark-mass profiles are shallower
than the true cluster profiles. The mass estimates based on
King-profile fits shown in panel (d) are significantly better than the
$\beta$-fit masses, but they are still systematically smaller than the
true masses by a factor of $\sim0.8$.

We have seen above that the low-biased $\beta$-fit mass estimates are
mainly caused by the fact that the $\beta$ fits to the X-ray emission
profiles are restricted to a range of radii within which the profiles
have not reached their asymptotic fall-off yet. King-profile fits are
biased low mainly if the cluster gas is not yet fully thermalized, in
which case the X-ray temperatures are lower than in equilibrium. To
emphasize this point, we compare in Fig.~10 the distribution of the
ratio of kinetic and thermal energy of the X-ray gas for arc clusters
(panel a) and for the entire cluster sample (panel b).

\begfigside{fig10.tps}
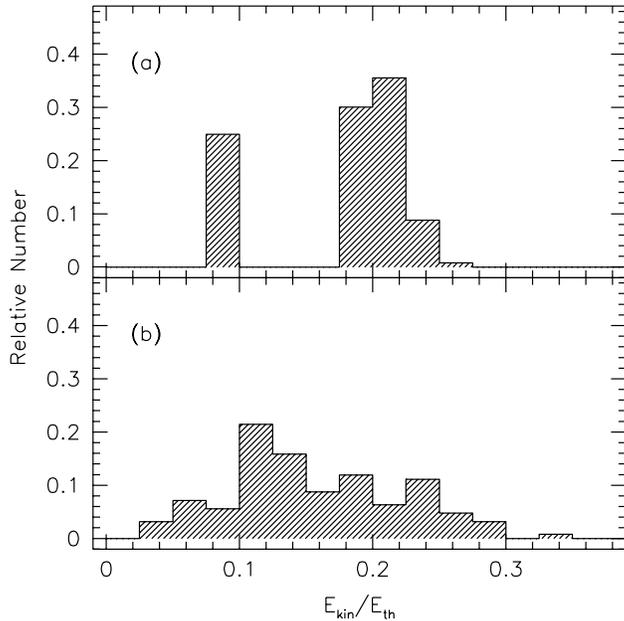
\figure{10}{Distribution of the ratio between kinetic and thermal
energy of the intracluster gas for arc clusters (panel a) and for the
entire cluster sample (panel b). On average, the arc clusters are less
thermalized, i.e., for them a larger fraction of the energy of the gas
is in bulk flows.}
\endfig

Figure 10 shows that the two distributions are significantly
different. The significance level of a Kolmogorov-Smirnov test for a
difference between the two distributions is 97 per cent. On average,
the fraction of non-thermalized energy of the intracluster gas is
higher in arc clusters than in the entire cluster sample, and the
distribution of $E_{\rm kin}/E_{\rm th}$ peaks at $\sim0.2$ for arc
clusters. The X-ray gas temperatures in arc clusters, and hence their
King-profile mass estimates, are thus too low by $\sim20$ per cent on
average, which is reflected by the histograms in panels (a) and (d) of
Fig.~9. The higher-than-average fraction of kinetic relative to
thermal energy indicates that bulk flows are prevalent in arc
clusters. This corresponds to the earlier finding (Bartelmann et
al. 1995) that substructure in clusters significantly increases their
ability to form arcs because they have a stronger tidal field than
spherically symmetric clusters and they produce caustic curves with a
larger number of cusp points. Substructure indicates ongoing merger
events, and during merger events the gas of the merging cluster
subclumps is not in thermal equilibrium.

If such clusters which have not yet reached their equilibrium state
are preferentially selected by strong-lensing events, we may suspect
that arc clusters are not necessarily those clusters with the highest
X-ray luminosities. The equilibrium luminosity of a cluster is
determined by the square of the gas density in the cluster core. While
clusters are still in the process of formation or merging from
subclumps, their X-ray luminosity should be lower than after
virialization. We therefore investigate the contribution of clusters
to the arc optical depth as a function of their X-ray luminosity
$d\tau/dL$. The results are shown in Fig.~11. There, we plot in panel
(b) the differential optical depth contributed by all clusters with
luminosity within $dL$ of $L$, and in panel (a) the same quantity
divided by the number of clusters in the corresponding luminosity
interval. Panel (a) of Fig.~11 therefore shows the contribution to the
optical depth per cluster.

\begfigside{fig11.tps}
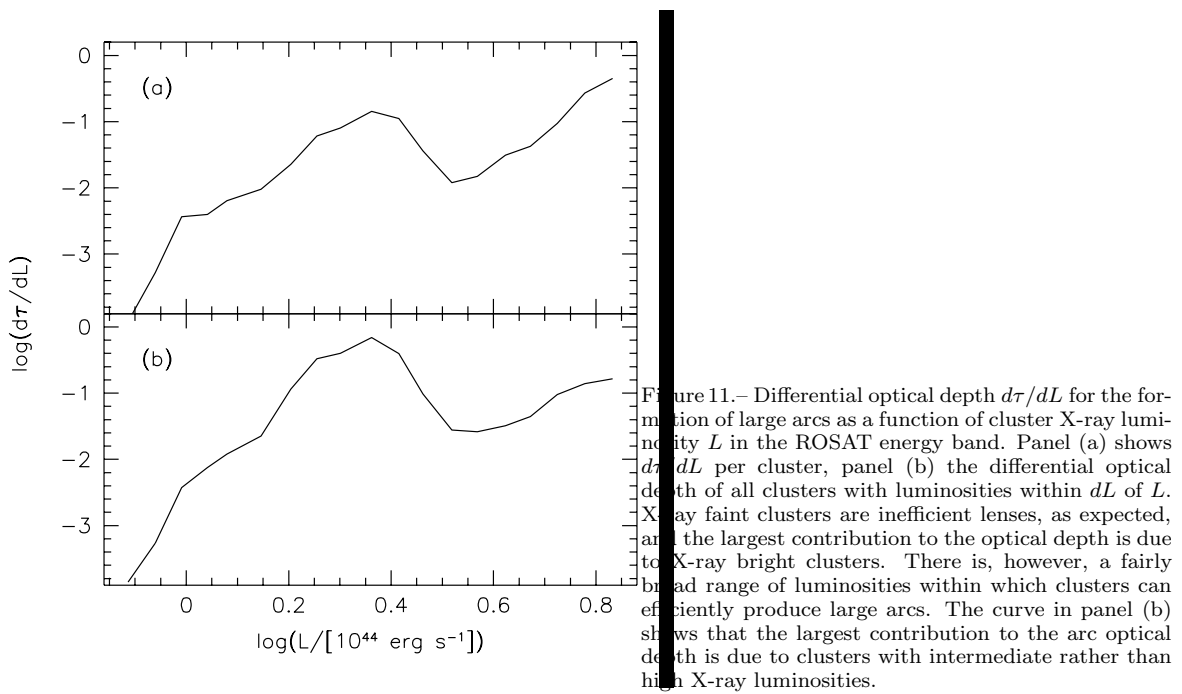
\figure{11}{Differential optical depth $d\tau/dL$ for the formation of
large arcs as a function of cluster X-ray luminosity $L$ in the ROSAT
energy band. Panel (a) shows $d\tau/dL$ per cluster, panel (b) the
differential optical depth of all clusters with luminosities within
$dL$ of $L$. X-ray faint clusters are inefficient lenses, as expected,
and the largest contribution to the optical depth is due to X-ray
bright clusters. There is, however, a fairly broad range of
luminosities within which clusters can efficiently produce large
arcs. The curve in panel (b) shows that the largest contribution to
the arc optical depth is due to clusters with intermediate rather than
high X-ray luminosities.}
\endfig

For low X-ray luminosities, the differential optical depth drops to
zero, as expected. Clusters which are not bright X-ray sources are not
massive or not concentrated enough in order to produce large arcs. The
brightest X-ray clusters also contribute most to the arc optical
depth, as seen in panel (b) of Fig.~11. There is, however, a fairly
broad range of cluster luminosities within which the contribution to
the optical depth per cluster is high, indicating that there is a
substantial contribution to the arc optical depth from clusters with
intermediate luminosities. If we consider cluster samples rather than
individual clusters, as in panel (b) of Fig.~11, it turns out that
clusters with intermediate X-ray luminosities, $1\times10^{44}\la
L/(\hbox{\rm erg s}^{-1})\la3\times10^{44}$, contribute the bulk of
the arc optical depth. The shape of the curves in Fig.~11 remains the
same if we split the cluster sample in halves and plot $d\tau/dL$
separately for each half. This indicates that the behaviour of the
curves is not due to noise.

The form of the curves in Fig.~11 can be qualitatively understood by
the following argument. Clusters with very high mass and therefore
high X-ray luminosity are efficient lenses with or without
substructure. Clusters with intermediate masses or luminosities are
efficient lenses only if they are substructured. The optical depth is
therefore composed of two contributions, one peaked contribution from
intermediate-mass clusters with substructure, and one monotonically
increasing contribution from high-mass clusters regardless of whether
they are substructured. Intermediate-mass clusters should be most
efficient lenses if and when they are in the process of formation
around the redshift where the critical surface-mass density is lowest,
i.e., at a redshift which is geometrically favored. For sources at
$z_{\rm s}=1$ which we assume throughout, $(d\ln\tau/dz)(z)$ sharply
peaks at $z\sim0.3$ and drops to half its peak value at $z\sim0.25$
and $z\sim0.4$ (Bartelmann et al. 1995). Intermediate-mass clusters
forming in the redshift interval $0.25\la z_{\rm f}\la0.4$ should
therefore contribute significantly to the optical depth because of
their substructure. The formation redshift of clusters is given as a
function of cluster mass by
$$
  z_{\rm f} = \left({M\over M_*}\right)^{-1/2}
\eqno(27)$$
(White 1994), where $M_*$ is the nonlinear mass today,
$$
  M_* = 6.3\times10^{13}\,M_\odot\,\sigma_8\,\Omega_0^{-0.7}
\eqno(28)$$
(White et al. 1993). The equilibrium X-ray luminosity scales with mass
as
$$
  L = 3\times10^{44}\left({M\over10^{15}\,M_\odot}\right)^{4/3}\,
  \left({\Omega_{\rm b}\over0.05}\right)^2
\eqno(29)$$
(Navarro et al. 1995a). These relations allow to translate the
redshift interval $0.25\la z_{\rm f}\la0.4$ to the luminosity interval
$1\times10^{44}\la L/(10^{44}\,\hbox{\rm erg s}^{-1})\la
3\times10^{44}$, which corresponds well to the luminosity range of the
peaks in Fig.~11. The maximum of the curves should be reached for
$z_{\rm f}\sim0.3$, corresponding to $L\sim1.9\times10^{44}$ erg
s$^{-1}$, which agrees well with the actual location of the peaks.

We can therefore conclude not only that the intrinsically brightest
clusters are efficient lenses, but also that there is a fairly broad
range of intermediate X-ray luminosities for which the contribution to
the optical depth reaches a maximum. This range is due to clusters
which form at redshifts for which the geometrical lens efficiency is
highest, because while they form their substructure increases the
optical depth to lensing. Only for low intrinsic luminosities does the
differential optical depth fall off. Figure 11 therefore demonstrates
that not the most luminous clusters are selected by lensing, but the
dynamically most active ones.
\titleb{Lensing Temperatures}
The presence of arcs in clusters imposes a constraint on the projected
cluster mass inside a cylinder with radius equal to the distance
between the arc and the cluster center. This mass constraint can be
converted to a temperature estimate. It was first pointed out by
Miralda-Escud\'e \& Babul (1995) that in two clusters for which X-ray
and lensing data were available the temperature required to explain
the lens effect is higher than the measured X-ray temperature by a
substantial amount.

We apply a comparable analysis to our numerically simulated clusters
here. This proceeds as follows. We fit $\beta$ models to the X-ray
surface-brightness profile of the clusters and obtain best-fit values
for $\beta$ and the core radius $r_{\rm c}$. Assuming that the gas is
isothermal, we obtain the cluster mass profile (15). Projecting this
along the line of sight, we obtain
$$
  M_\parallel(r) = {3\pi\over2}
  {\beta kT\,r_{\rm c}\over G\mu m_{\rm p}}\,
  {x^2\over\sqrt{1+x^2}}\;.
\eqno(30)$$
For the projected mass inside the arc radius we take the lensing mass
as determined before. If we assumed spherical symmetry instead, the
lensing mass estimate would systematically be biased high, as
mentioned before. Hence we get from eq.~(30)
$$
  kT_{\rm lens} = 
  {2G\mu m_{\rm p}M_{\rm lens}\over3\pi\beta r_{\rm c}}
  \,{\sqrt{1+x^2}\over x^2}\;.
\eqno(31)$$
The equation shows that $kT_{\rm lens}$ is proportional to
$\beta^{-1}$. A flatter gas profile needs to be hotter for fixed dark
binding mass. The lensing temperature should therefore
straightforwardly reflect the $\beta$ discrepancy. Since
$\beta\sim2/3$ on average, the lensing temperatures are expected to be
too high by a factor of $\sim1.5$ on average. We show in Fig.~12 the
distribution of the ratio between the lensing temperature and the
spectral X-ray temperature of the intracluster gas.

\begfigside{fig12.tps}
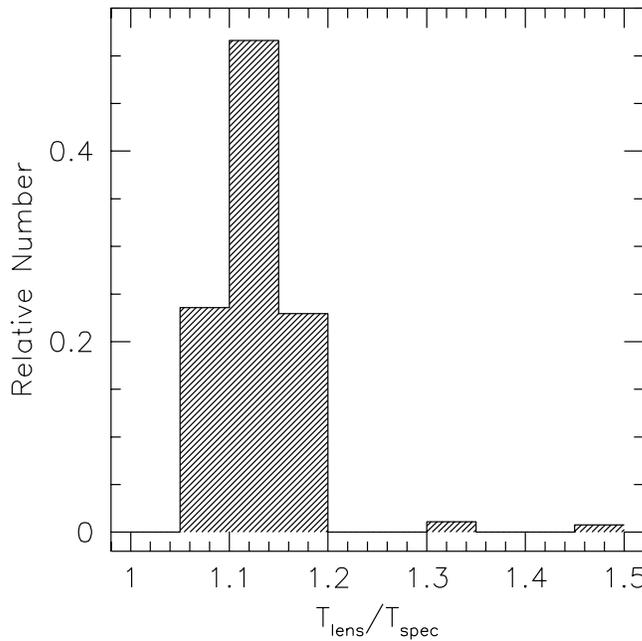
\figure{12}{Distribution of the ratio between lensing temperature
$T_{\rm lens}$ and X-ray spectral temperature \break $T_{\rm
spec}$. The distribution peaks at $\sim1.15$, showing that on average
the lensing temperature well reflects the temperature of the cluster
gas with only a weak bias towards higher values of $T_{\rm lens}$.}
\endfig

Contrary to this expectation, the lensing temperature is higher than
the spectral temperature by only $\sim15$ per cent. This result must,
however, be interpreted with caution. Equation (16) shows that the
dark-matter density profile derived from isothermal $\beta$ fits falls
off $\propto r^{-2}$ for large $r$, while the true density profile
asymptotically follows $r^{-3}$. When projected along the line of
sight, the shallower density profile contributes much more mass from
large radii than the steeper profile. The flatter the profile is, the
less projected mass is concentrated in the cluster center, and if
there is less mass in the center, the central gas temperature can be
lower. Clearly, the density profile (16) must cut off or steepen at
some large radius beyond which the assumption of isothermality breaks
down. In that case, $M_\parallel$ would be smaller than expected from
(30), hence $kT_{\rm lens}$ would be larger. An ambiguity then arises
as to where and how the isothermal mass profile should be cut
off. Thus there are two competing effects: (1) the low bias of the
$\beta$ values tends to increase $T_{\rm lens}$, while (2) the
projection along the line-of-sight of the flat isothermal mass profile
tends decrease $T_{\rm lens}$. Figure 12 demonstrates that the
systematically low $\beta$ values are almost compensated by the
projection of the isothermal mass profile along the line-of-sight. We
can therefore conclude that on average, the effect demonstrated by
Miralda-Escud\'e \& Babul (1995) does not occur in our sample, or in
other words that for an average cluster the lensing temperature should
well reflect the X-ray temperature although being somewhat too high.

However, we can identify several clusters in our sample for which
$kT_{\rm lens}$ is higher by up to a factor two than the X-ray
temperature. Given the potential importance of this discrepancy, these
clusters deserve closer inspection.

We plot in Fig.~13 the distribution of radial velocities of dark
cluster particles in a cylinder of $1\,h^{-1}$ Mpc radius around the
line-of-sight towards the three cluster models which show the largest
discrepancy between $T_{\rm lens}$ and the X-ray spectral temperature
$T_{\rm spec}$.

\begfigside{fig13.tps}
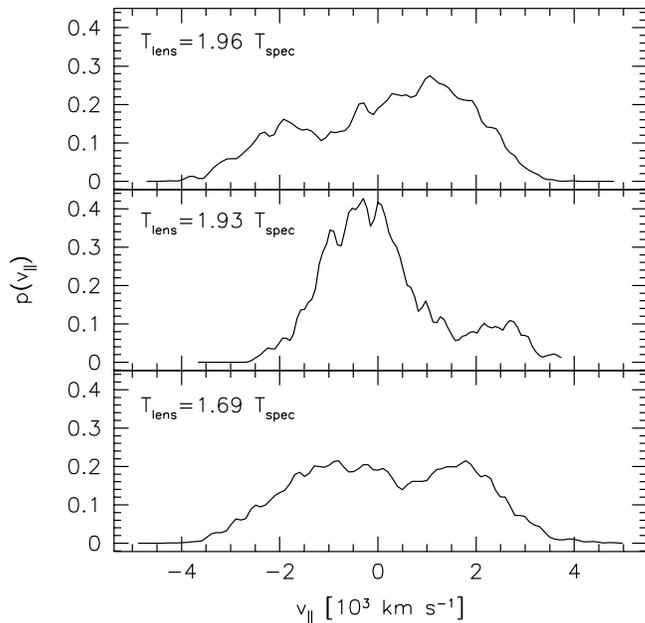
\figure{13}{Distribution of radial velocities of dark cluster
particles for the three numerical cluster models in our sample which
show the largest discrepancies between $T_{\rm lens}$ and $T_{\rm
spec}$. The distributions are centered, and $v_\parallel$ is given in
$10^3$ km s$^{-1}$. All three distributions have secondary peaks or
are bimodal, indicating infall along the line-of-sight.}
\endfig

The curves in Fig.~13 show that the radial velocity distributions for
these three clusters have secondary peaks or are bimodal. This
indicates infall of cluster subclumps along the line-of-sight. In
these cases, the matter density of all cluster clumps projected along
the line-of-sight is larger than expected in a spherically symmetric
model, and at the same time the X-ray temperature and luminosity are
lower than expected if the total cluster matter was spherically
symmetric. Therefore, although statistically unimportant, projection
effects do play a role in some individual clusters for which the
lensing- and X-ray temperatures are found to be discrepant.

To clarify this further, we plot in Fig.~14 the relation between the
line-of-sight velocity dispersion $\sigma_\parallel$, divided by the
average velocity dispersion $\langle\sigma\rangle$, and the ratio
between the lensing- and X-ray temperatures $T_{\rm lens}/T_{\rm
spec}$. For clusters with $\sigma_\parallel/\langle\sigma\rangle>1$,
the line-of-sight points along a direction into which the velocity
ellipsoid of the dark cluster particles is elongated.

\begfigside{fig14.tps}
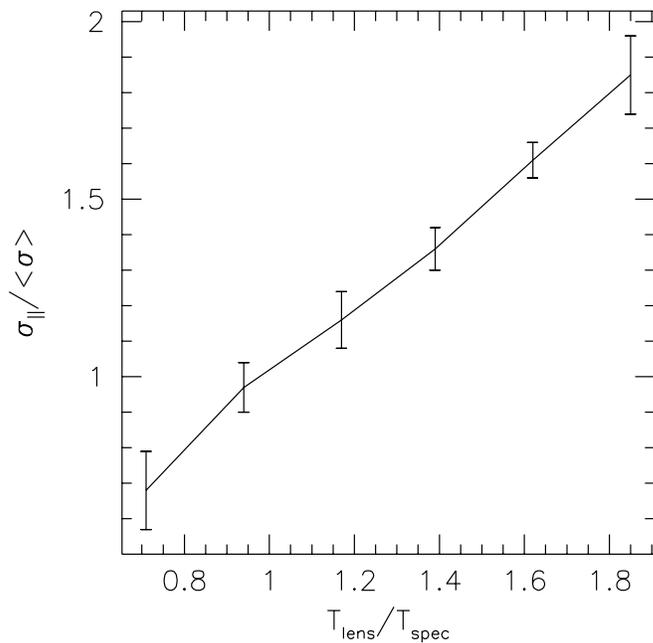
\figure{14}{Relation between the velocity dispersion along the
line-of-sight $\sigma_\parallel$, in units of the average velocity
dispersion $\langle\sigma\rangle$, and the ratio between lensing
temperature $T_{\rm lens}$ and X-ray spectral temperature $T_{\rm
spec}$. Clusters whose velocity ellipsoid is elongated along the
line-of-sight show a large discrepancy between lensing- and spectral
temperature, indicating that this temperature discrepancy can be
attributed to projection effects.}
\endfig

Figure 14 shows that there is an evident and strong correlation
between the velocity dispersion along the line-of-sight and the
lensing temperature, confirming that high values of $T_{\rm lens}$
relative to $T_{\rm spec}$ occur in such clusters which exhibit
structure along the line-of-sight. The clusters analyzed by
Miralda-Escud\'e \& Babul (1995) indeed show structure in velocity
space or indications of ongoing merging (Teague, Carter, \& Gray 1990;
Markevitch et al. 1996; Daines et al. 1996), in agreement with our
findings.
\titlea{Summary and Discussion}
We have performed gas-dynamical simulations of galaxy clusters in
order to compare their X-ray emission with their strong-lensing
properties. The simulations were done within the CDM cosmogony. The
density-perturbation power spectrum was normalized to $\sigma_8=1$,
and the cosmological parameters $\Omega_0=1$, $\Lambda_0=0$, and
$H_0=50$ km s$^{-1}$ Mpc$^{-1}$ were adopted. The baryon fraction was
chosen to be $\Omega_{\rm b}=0.05$. Clusters were identified in a
large simulation volume and later re-simulated at much increased
spatial resolution, taking the tidal effects of surrounding matter
into account. We selected a sample of thirteen clusters with masses
$6\times10^{14}\le M_{\rm 500}/M_\odot\le4\times10^{15}$ and examined
them at redshifts between zero and unity.

We calculated the X-ray emission of the cluster gas due to thermal
bremsstrahlung and analyzed it imitating the imaging and spectral
properties of current X-ray telescopes. Images were taken with the
spatial resolution of the ROSAT HRI, and spectra were taken with the
spatial and spectral resolution of the ASCA SIS. All X-ray analyses
were done after discretizing the X-ray emission into photons and after
adding background noise. In particular, we derived the temperature of
the intracluster gas by fitting the thermal bremsstrahlung spectrum to
the synthetic cluster spectra, rather than taking the temperature as
given from the numerical simulations. The temperature determination
was accurate to $\sim15$ per cent on average. There is no systematic
trend with temperature in the accuracy of the temperature
determination. A total of 378 cluster images were analyzed.

We then derived mass estimates for the X-ray clusters. The first and
most widely used technique to do so is to fit the $\beta$ model to the
azimuthally averaged cluster emission profiles. We find that the
best-fitting $\beta$ values are $\sim2/3$ on average, however with a
fairly large scatter. Values of $\beta\sim1$ are expected because the
the cluster gas in our simulations is in equilibrium with the
dark-matter particles of the clusters within the radial range where
the fits can be done. Reducing the background yields significantly
higher $\beta$ values. This is in agreement with the earlier finding
by Navarro et al. (1995a) that the systematically low $\beta$ values
are due to the rather limited range of radii accessible through X-ray
observations. Where the X-ray emission falls below the background, the
emission profiles usually have not reached their asymptotic slopes
yet, and hence the best-fit $\beta$ values are systematic
underestimates.

The density profiles of the clusters in our simulations are well
reproduced by the profile found recently by Navarro et al. (1995b),
which falls off $\propto r^{-3}$ asymptotically and approaches the
center $\propto r^{-1}$. Isothermal mass profiles with a flat core
derived from $\beta$ fits are therefore too shallow in the outer and
inner parts of the clusters. Hence the accuracy of the $\beta$-fit
mass estimates changes with radius. At the maximum radius accessible
to X-ray observations, where the X-ray emission profile falls below
the X-ray background, the $\beta$-fit masses reach $\sim60$ per cent
of the true mass for the average value $\beta\sim2/3$.

Since $\beta$ turned out to be systematically too low, we investigated
fitting King profiles to the cluster emission rather than $\beta$
models. This is equivalent to fitting the profiles constraining
$\beta$ to unity. Over the relevant range of radii, the density
profile of the dark matter in the numerically simulated clusters can
well be described by a King profile, except at the very center. It
turns out that the resulting mass estimates are substantially more
accurate than the $\beta$-fit mass estimates despite the reduced
degree of freedom. Yet, the mass estimates are systematically low by
$\sim10\ldots20$ per cent. We find that this remaining bias is due to
bulk flows in the intracluster gas. Bulk flows indicate that the gas
is not fully thermalized, so that the gas temperature is lower than it
should be if the gas was completely thermalized. Observationally, they
may be detectable by peaks in the temperature distribution of X-ray
clusters.

We then investigated the properties of such X-ray clusters which are
able to produce large arcs. We weighted cluster properties such as
their mass and the relative kinetic energy of the cluster gas with the
optical depth to the formation of large arcs. The distribution of
lensing mass estimates relative to the distribution of true cluster
masses shows that projection effects are statistically unimportant in
a sample of clusters. It is however crucial not to estimate lensing
masses based on the assumption of axially symmetric lenses because in
that case the lensing mass estimates are substantially too
high. $\beta$-fit estimates of the cluster mass inside the arc radius
are systematically lower by a factor of $1.5\ldots2$ than the true
cluster masses, reflecting again the facts that (1) best-fit $\beta$
values are systematically too low and (2) the isothermal $\beta$-fit
mass profile is flatter than the true density profile. The mass
estimates based on the King profile are significantly more accurate,
although still biased low by $\sim20$ per cent.

The reason for this latter bias is that the fraction of
non-thermalized, kinetic energy in the intracluster gas is
significantly higher in the arc clusters than in the entire cluster
sample. In most arc clusters, the kinetic energy of the gas reaches
$\sim20$ per cent of the thermal energy, indicating that the gas
temperature is only $\sim80$ per cent of what it should be if the
intracluster gas was completely thermalized. The X-ray temperature in
lensing clusters is therefore systematically too low. This reiterates
the point that strong lensing selects such clusters which show
substructure. Ongoing merging of substructure gives rise to bulk flows
in the intracluster gas. We also show that the bulk of the optical
depth for large arcs is contributed by clusters with intermediate
rather than high X-ray luminosity.

Finally, we perform an analysis similar to that done by
Miralda-Escud\'e \& Babul (1995), where they compared the cluster
temperature demanded by the presence of large arcs on the basis of
hydrostatic equilibrium to the measured X-ray temperature and found
that for two of the three clusters they studied the lensing
temperature was higher by a factor of $\sim2$ than the X-ray
temperature. On average, we find that the lensing temperature in our
clusters well reproduces the X-ray spectral temperature with only a
small high bias of $\sim15$ per cent. There are, however, clusters in
our sample for which the lensing temperature is indeed higher by about
a factor of two than the X-ray temperature. We find that those
particular clusters show structure in velocity space, indicating
ongoing merging along the line-of-sight. This gives rise to projection
effects which increase the lensing mass and simultaneously reduce the
X-ray mass.

We can therefore summarize our major findings in the following way.
\item{1.} The $\beta$ model yields systematically low cluster mass
estimates because within the limited radial range accessible to X-ray
observations the profile is usually shallower than it is
asymptotically.
\item{2.} Fitting a King-profile to the X-ray emission substantially
improves the mass estimates. These are still systematically too low by
$\sim10$ per cent on average.
\item{3.} The presence of bulk flows in the intracluster gas reduces
the X-ray temperature by an amount proportional to the fraction of
kinetic energy in the bulk flows. This accounts for the systematic
bias remaining in the King-profile mass estimates. Bulk flows are
accompanied by shocks in the intracluster gas which lead to
significant local peaks in the temperature distribution of the
clusters.
\item{4.} Strong lensing selects clusters which are dynamically more
active than the average. The distribution of the relative kinetic
energy of the intracluster gas peaks at significantly higher values in
arc clusters compared to the entire cluster sample, indicating the
prevalence of bulk flows and thus of ongoing merging in arc clusters.
\item{5.} The bulk of the optical depth to strong lensing is
contributed by clusters with intermediate rather than high X-ray
luminosities.
\item{6.} Cluster temperatures derived from the presence of large arcs
in clusters are generally in good agreement with the X-ray
temperatures of the same clusters. Such clusters where the lensing
temperature is significantly higher than the X-ray temperature
routinely show structure in velocity space which indicates that the
temperature discrepancy is due to projection of multiple cluster
subclumps along the line-of-sight.

\acknow{~We wish to thank Simon White and Peter Schneider for many
useful comments. This work was supported in part by the
Sonderforschungsbereich SFB 375-95 of the Deutsche
Forschungsgemeinschaft.}
\begref{References}
\ref Bahcall, N.A., \& Lubin, L.M. 1994, ApJ, 426, 513
\ref Bartelmann, M., \& Weiss, A. 1994, A\&A, 287, 1
\ref Bartelmann, M., Steinmetz, M., \& Weiss, A. 1995, A\&A 297,
1
\ref Bartelmann, M. 1995, A\&A, 299, 11
\ref Bartelmann, M., Narayan, R., Seitz, S., \& Schneider,
P. 1996, ApJ Letters, in press
\ref Binney, J., \& Tremaine, S. 1987, Galactic Dynamics
(Princeton: University Press)
\ref Bonnet, H., Mellier, Y., \& Fort, B. 1994, ApJ, 427, L83
\ref Briel, U.G., \& Henry, J.P. 1994, Nat, 372, 439
\ref Broadhurst, T.J., Taylor, A.N., \& Peacock, J.A. 1995, ApJ,
438, 49
\ref Bryan G.L., Cen R., Norman M.L., Ostriker J.P., \& Stone
J.M. 1994, ApJ, 428, 405
\ref Cavaliere, A., \& Fusco-Femiano, R. 1976, A\&A, 49, 137
\ref Cen R., \& Ostriker J.P. 1994, ApJ 429, 4
\ref Cole, S., \& Lacey, C. 1995, SISSA preprint
astro-ph/9510147
\ref Daines, S., Jones, C., Forman, W. \& Tyson, J.A. 1996, ApJ,
submitted
\ref David, L.P., Harnden Jr., F.R., Kearns, K.E., \& Zombeck,
M.V. 1993, U.S. ROSAT Science Data Center / Smithsonian Astrophysical
Observatory
\ref Edge, A.C., \& Stewart, G.C. 1991, MNRAS, 252, 428
\ref Evrard, A.E. 1990, ApJ, 363, 349
\ref Evrard, A.E., Metzler, C.A., \& Navarro, J.F. 1995, SISSA
preprint astro-ph/9510058
\ref Gingold, R.A., \& Monaghan, J.J. 1977, MNRAS, 181, 375
\ref Henriksen, M.J., \& Mushotzky, R. 1985, ApJ, 292, 441
\ref Henry, J.P., \& Briel, U.G. 1995, ApJ, 443, L9
\ref Jones, C., \& Forman, W. 1984, ApJ, 276, 38
\ref Kaiser, N., \& Squires, G. 1993, ApJ, 404, 441
\ref Katz, N., \& White, S.D.M. 1993, ApJ, 412, 455
\ref King, I.R. 1966, AJ, 71, 64
\ref Lubin, L.M., \& Bahcall, N.A. 1993, ApJ, 415, L17
\ref Lucy, L. 1977, AJ, 82, 1013
\ref Markevitch, M., Mushotzky, R., Inoue, H., Yamashita, K.,
Furuzawa, A., \& Tawara, Y. 1996, ApJ, 456, 437
\ref Miralda-Escud\'e, J., \& Babul, A. 1995, ApJ, 449, 18
\ref Monaghan, J.J. 1992, ARA\&A, 30, 543
\ref Navarro, J.F., Frenk, C.S., \& White, S.D.M. 1995a, MNRAS,
275, 720
\ref Navarro, J.F., Frenk, C.S., \& White, S.D.M. 1995b, MNRAS,
in press
\ref Sarazin, C. 1986, Rev. Mod. Phys., 58, 1
\ref Schindler, S., \& M\"uller, E. 1993, A\&A, 272, 137
\ref Schneider, P., Ehlers, J., \& Falco, E.E. 1992,
Gravitational Lenses (Heidelberg: Springer)
\ref Seitz, C., \& Schneider, P. 1995, A\&A, 297, 287
\ref Seitz, S., \& Schneider, P. 1996, A\&A, 305, 383
\ref Serlemitsos, P.J., Jalota, L., Soong, Y., Kunieda, H. et
al. 1995, PASJ, 47, 105
\ref Snowden, S.L., Freyberg, M.J., Plucinsky, P.P., Schmitt,
J.H.M.M. et al. 1995, ApJ, 454, 643
\ref Squires, G., \& Kaiser, N. 1995, SISSA preprint
astro-ph/9512094
\ref Steinmetz, M. 1996, MNRAS, 278, 1005
\ref Takahashi, T., Markevitch, M., Fukazawa, Y., Ikebe, Y.,
Ishisaki, Y., Kikuchi, K., Makishima, K., \& Tawara, Y. 1995, ASCA
Newsletter, No. 3 (NASA/GSFC)
\ref Tanaka, Y., Inoue, H., \& Holt, S.S. 1994,
Pub. Astr. Soc. Japan, 46, L37
\ref Teague, P.F., Carter, D., \& Gray, P.M. 1990, ApJS, 72, 715
\ref Thomas, P.A., \& Couchman, H.M.P. 1992, MNRAS, 257, 11
\ref White, D.A., \& Fabian, A.C. 1995, MNRAS, 273, 72
\ref White, S.D.M., 1994, Les Houches Lectures
\ref White, S.D.M., Efstathiou, G., \& Frenk, C.S. 1993, MNRAS,
262, 1023
\ref White, S.D.M., Navarro, J.F., Evrard, A.E., \& Frenk,
C.S. 1993, Nat, 366, 429
\ref Wilson, G., Cole, S., \& Frenk, C. 1995, SISSA preprint
astro-ph/9601102
\ref Zwicky, F. 1933, Helv. Phys. Acta, 6, 110
\endref
\end